\documentclass[twocolumn,showpacs,preprintnumbers,amsmath,amssymb]{revtex4}
\usepackage{epsfig}

\def\lsim{\mathrel{\rlap{\lower4pt\hbox{\hskip1pt$\sim$}}
    \raise1pt\hbox{$<$}}}                
\def\gsim{\mathrel{\rlap{\lower4pt\hbox{\hskip1pt$\sim$}}
    \raise1pt\hbox{$>$}}}                

\begin{document}

\title{Non-equilibrium work fluctuations for oscillators in non-Markovian baths}

\author{Trieu Mai$^1$ and Abhishek Dhar$^2$}
\affiliation{$^1$Department of Physics, University of California,
Santa Cruz, California 95064, USA}
\affiliation{$^2$Raman Research Institute, Bangalore 560080, India}

\date{\today}

\begin{abstract}
We study work fluctuation theorems for oscillators in non-Markovian
heat baths. By calculating the work distribution function for a
harmonic oscillator with motion described by the generalized Langevin
equation, the Jarzynski equality (JE), transient fluctuation theorem
(TFT), and Crooks' theorem (CT) are shown to be exact. In addition
to this derivation, numerical simulations of anharmonic oscillators
indicate that the validity of these nonequilibrium theorems do not
depend on the memory of the bath. We find that the JE and the CT
are valid under many oscillator potentials and driving forces whereas
the TFT fails when the driving force is asymmetric in time {\it
and} the potential is asymmetric in position.
\end{abstract}


\maketitle 

\section{Introduction}

Fluctuation theorems (FTs) which describe properties of the distribution
of various nonequilibrium quantities, such as work and entropy,
have been developed over the past
decade~\cite{phystod,dynsys,cohen,jarzynski,jarzynski2,crooks,ND}. Unlike most 
other relations in nonequilibrium statistical mechanics, remarkably,
the FTs are applicable to systems driven arbitrarily far from
equilibrium. 

In particular, consider a classical finite-sized system in contact
with a heat bath at temperature $T$ and driven by some generalized
force $\lambda$ (e.g., volume, magnetic field). In equilibrium, the
phase space distribution of the system is described by a well-defined
statistical mechanical ensemble. The free energy $F$ can, in
principle, be calculated and is a function of these parameters,
i.e. $F=F(T,\lambda)$. At some time $t$, say $t=0$, $\lambda$ is
varied via a fixed path $\lambda(t)$ to a later time $t=\tau$.
If this process is carried out reversibly, the work done $W$ is
simply the free energy difference $\Delta
F=F(T,\lambda_\tau)-F(T,\lambda_0)$. More generically, the process
is irreversible and the second law gives the well-known inequality
\begin{equation}
W \geq \Delta F.
\label{eq:inequality}
\end{equation}

An ensemble of such processes would yield the work probability
distribution $P(W)$. The finite width of $P(W)$ is due to two
stochastic sources: (1) The system configuration when the force is
first initiated at $t=0$ is drawn from the equilibrium ensemble and
(2) the path the system travels (in phase space) during the driving
process is not deterministic due to the coupling with the heat bath.
We study $P(W)$ in the light of three closely related theorems
described below, the Jarzynski equality (JE)~\cite{jarzynski}, the
transient fluctuation theorem (TFT)~\cite{crooks,ND,dynsys}, and
Crooks' theorem (CT)~\cite{crooks,ND}.

The powerful nonequilibrium work relation due to Jarzynksi~\cite{jarzynski}
allows one to exactly obtain equilibrium information (the free
energy difference) from measurements of nonequilibrium processes.
The JE states
\begin{equation}
<e^{-\beta W}>=e^{-\beta \Delta F},
\label{eq:JE}
\end{equation}
where $\beta$ is the inverse temperature (with Boltzmann's constant
set to unity) and the average is over the distribution function
$P(W)$ described above. The {\it equality} allows for the computation
of $\Delta F$ even when the driving process is not adiabatic. This
is to be compared with the {\it inequality} of Eq.(\ref{eq:inequality}).
The JE has been shown for Hamiltonian systems~\cite{jarzynski} and
Markovian stochastic systems~\cite{jarzynski2,crooks,ND}.

Fluctuation theorems have been found for a wide class of systems
and various nonequilibrium quantities, including work, heat, and
entropy production. They are also generally divided into steady
state~\cite{cohen} and transient theorems~\cite{crooks,ND,dynsys}.
In this paper, we restrict our discussion to the transient fluctuation
theorem for the mechanical work done. The TFT relates the ratio of
probability distributions for the production of positive work to
the production of negative work, 
\begin{equation} 
\frac{P(+W)}{P(-W)}
= e^{+\beta W}.  
\label{eq:TFT} 
\end{equation} 
Similar to the JE,
the TFT has been derived in driven deterministic systems~\cite{dynsys},
and Markovian stochastic systems~\cite{crooks,ND,markov1,markov2,markov3}.

Crooks~\cite{crooks} connected the JE and the TFT by using a relation
very similar to Eq.(\ref{eq:TFT}), which we refer to as the Crooks'
theorem,
\begin{equation}
\frac{P_F(+W)}{P_R(-W)} = e^{+\beta W},
\label{eq:CT}
\end{equation}
where $P_F(+W)$ is the probability distribution for work done as
described in the above scenario and $P_R(-W)$ is the probability
distribution for negative work done in a time-reversed driving
process. The similarity between Eq.(\ref{eq:TFT}) and Eq.(\ref{eq:CT})
is evident and the two theorems are, in fact, equivalent for a large
class of systems. However, under certain asymmetries in the potential
and the driving force, differences between the TFT and CT arise. 
In Sec. IV, we show one such scenario.

Unlike some other nonequilibrium quantities (e.g. heat and entropy), the classical
mechanical work is an easily defined quantity, $W = \int_0^\tau
dt f(t) \dot{x}(t)$. The original formulation of the JE~\cite{jarzynski},
however, relies on the generalized work, $W_J = -\int_0^\tau dt
\dot{f}(t) x(t)$. As discussed in Refs.~\cite{ND,polymer}, these
two differ by the boundary conditions $f(\tau)x(\tau)-f(0)x(0)$ and
care must be taken in defining the work in the fluctuation theorems.

The (space and time) extensiveness of $W$ (or $W_J$) demonstrates
that second law ``violations'' become exponentially unlikely for
thermodynamic systems; these violations only become observably
probable for microscopic systems. Beyond purely theoretical
interests, the recent investigations of molecular motors and
nano-mechanical devices demonstrate the practical importance for
understanding these universal nonequilibrium theorems, especially
under increasingly realistic scenarios. Technological accessibility
of microscopic systems has opened up experimental study and
verification of the above theorems in various
systems~\cite{experiment,liphardt}.

In this paper, we focus on the JE, TFT, and CT for single harmonic
and anharmonic oscillators. Using second order Langevin dynamics
of a single oscillator, the work fluctuation theorems have been
studied in Ref.~\cite{harmonic}. The work distribution function in
harmonic polymer chains has been studied by one of us~\cite{polymer}.
Both these papers and the derivations of the FTs for stochastic
systems~\cite{crooks,ND,markov1,markov2,markov3} utilize Markovian
dynamics. Here, we study the FTs for oscillators coupled to
non-Markovian baths. This extension is motivated
by the intuition that there must be some memory to real heat baths
and that all realistic models of heat baths, either classical~\cite{classicbath}
or quantum mechanical~\cite{quantumbath}, are non-Markovian. Through
analytic expressions for the harmonic oscillator and numerical
simulations of anharmonic oscillators, we demonstrate that the
memory of the bath does not affect the validity of the FTs.

In the next section, we briefly describe the model and the generalized
Langevin dynamics. Sec. III contains explicit derivations of the
JE, TFT, and CT in the harmonic limit where analytic calculations
are possible. We show numerical results for the nonlinear oscillators
in Sec. IV and summarize our results in the last section.

\section{the Model}

We model the system as a unit mass particle in a one-dimensional
potential with dynamics governed by the generalized Langevin
equation~\cite{genlangevin},
\begin{equation}
\ddot{x}(t) = -\frac{dV(x)}{dx} + f(t) - \int_0^t dt^\prime 
\gamma(t-t^\prime)\dot{x}(t^\prime) + \eta(t),
\label{eq:langevin}
\end{equation}
where $V(x)$ is a conservative oscillator potential and$f(t)$ is
an externally determined time-dependent driving force. The $\eta(t)$
and $\gamma(t)$ terms represent Gaussian noise and damping,
respectively, and must be related through the fluctuation-dissipation
theorem,
\begin{equation}
<\eta(t)> = 0, \qquad <\eta(t)\eta(t^\prime)> = T\gamma(t-t^\prime),
\label{eq:FD}
\end{equation}
where $T$ is the temperature of the bath. In the white Gaussian
noise limit, $\gamma(t)$ is proportional to a Dirac $\delta$-function
and the more familiar ``Markovian'' Langevin equation is recovered.
We mention that the damping is even in time $\gamma(t)=\gamma(-t)$,
an important property to be used in the next section.

We use a potential $V(x)$ of the form 
\begin{equation}
V(x) = \frac{\omega_o^2x^2}{2}+\frac{k_3x^3}{3}+\frac{k_4x^4}{4}
\label{eq:pot}
\end{equation}
which can represent a truncation of the Taylor expansion of some
complicated potential. We restrict ourselves to bounded potentials.
$x$ can represent the spatial position, an angular variable as in
Ref.~\cite{harmonic}, or a generalized coordinate. From this potential
and the driving force $f(t)$ conjugate to $x$, the Hamiltonian
$\mathcal{H}$ is clearly $\mathcal{H}(t)=\dot{x}^2/2+V(x)-f(t)x$.
If the force is time-independent, the system would eventually reach
equilibrium and the free energy $F$ can be simply calculated by $F=-T
ln(\mathcal{Z})$, where $\mathcal{Z}$ is the partition function.
In this ensemble, we recall that the Jarzynski work $W_J$ is not generally
equal to the real mechanical work $W$~\cite{ND,polymer}. In the
next section, we derive the work distribution function for the harmonic
oscillator, i.e. $k_3=k_4=0$.

\section{The Harmonic Oscillator}

At low temperatures, many potentials can be well approximated by
the harmonic potential. Harmonic oscillators also have the practical
virtue that a formal solution to Eq.(\ref{eq:langevin}) exists. We
use this formal solution to derive the distribution functions for
the Jarzynski generalized work $W_J$ and the real mechanical work
$W$. The work distribution functions are then used to verify the
JE and the TFT. (The TFT and CT are equivalent for the harmonic
oscillator.) The analytic expressions in this section are analogous
to the expressions from Ref.~\cite{polymer} with three main
differences. The primary difference is that the noise is colored
here. Second, we use the full second order Langevin equation instead
of taking the strongly overdamped limit. The last difference is
that we only study single harmonic oscillators instead of harmonic
chains. Our results should generalize to chains, which are more
applicable to polymer stretching experiments~\cite{liphardt}, though
we do not discuss this generalization further.

Equilibrium quantities are easily evaluated for the harmonic
oscillator. Under constant driving, the free energy is
\begin{equation}
F(T,f) = -\frac{f^2}{2\omega_o^2}-{T} ln\frac{2\pi T}{\omega_o}. 
\end{equation}
Since only the free energy difference $\Delta F$ appears in the JE,
only the first term on the right hand side is relevant. In this 
equilibrium ensemble, the averages for the initial position $x_o$ 
and velocity $v_o$ and their variances are
\begin{eqnarray}
<x_o> = f/\omega_o^2,&& \sigma_{x_o}^2 = T/ \omega_o^2, \nonumber \\
<v_o> = 0,&& \sigma_{v_o}^2 = T,
\end{eqnarray}
where $\sigma^2_A = <A^2>-<A>^2$ for any quantity $A$.

The formal solution to Eq.(\ref{eq:langevin}) for a harmonic potential
is (for $t>0$):
\begin{equation}
x(t)=H(t)x_o+G(t)v_o +\int_0^t dt^\prime G(t-t^\prime)[f(t^\prime)+
\eta(t^\prime)],
\label{eq:sol}
\end{equation}
where $H(t)$ and $G(t)$ are the homogeneous solutions with properties
$H(0)=\dot{G}(0)=1$ and $\dot{H}(0)=G(0)=0$. For $t<0$ we define
$H(t)=H(-t)$ and $G(t)=-G(-t)$.
The stochastic terms
are $x_o$, $v_o$, and $\eta$. Using the definition of $W_J$ ($W$),
we see that $W_J$ ($W$) is proportional to $x$ ($\dot{x}$) and is,
therefore, a linear combination of the stochastic terms. Thus, in
the harmonic limit, $W_J$ is Gaussian distributed and it is sufficient
to calculate the mean $<W_J>$ and variance $\sigma_J^2 $. Obviously,
$W$ is Gaussian as well and its distribution function is
\begin{equation}
P(W) = \frac{1}{\sqrt{2\pi\sigma^2}}e^{-\frac{(W-<W>)^2}{2\sigma^2}}.
\end{equation}
For such Gaussian processes, the TFT is satisfied if $<W>=\beta\sigma^2/2$
and the JE is satisfied if $<W_J>=\Delta F+\beta\sigma^2/2$.

In order to compare the means and variances of the work distributions, we first
derive identities for the Green's functions $H(t)$ and $G(t)$ in
Eq.(\ref{eq:sol}). The Laplace transform of Eq.(\ref{eq:langevin})
is
\begin{equation}
s^2\tilde{x}(s)-sx_o-v_o+\tilde{\gamma}(s)[s\tilde{x}(s)-x_o]+
\omega_o^2\tilde{x}(s) = 0,
\label{eq:laplace}
\end{equation}
where $\tilde{x}(s)=\int_0^\infty dt x(t)e^{-st}$ is the standard
definition of the Laplace transform. The Green's functions, $H(t)$
and $G(t)$, must also satisfy equations analogous to Eq.(\ref{eq:laplace})
and simplify due to their initial conditions. Solving these algebraic
expressions gives
\begin{eqnarray}
\tilde{H}(s) & = & \frac{\tilde{\gamma}+s}{s^2+s\tilde{\gamma}+
\omega_o^2}, \nonumber \\
\tilde{G}(s) & = & \frac{1}{s^2+s\tilde{\gamma}+\omega_o^2} .
\end{eqnarray}
Some manipulations and the inverse transform reveal identities
between the two Green's functions in Laplace and real space,
\begin{eqnarray}
s\tilde{H}(s) & = & 1-\omega_o^2 \tilde{G}(s), \nonumber \\
s\tilde{G}(s) & = & \tilde{H}(s) - \tilde{\gamma}(s)\tilde{G}(s), 
\nonumber \\
\dot{H}(t) & = & -\omega_o^2G(t), \nonumber \\
\dot{G}(t) & = & H(t) - \int_0^t dt^\prime \gamma(t-t^\prime)G(t^\prime).
\label{eq:ids} 
\end{eqnarray}
For white noise, where $H(t)$ and $G(t)$ are well-known, we confirm
that Eqs.(\ref{eq:ids}) are correct.

We plug Eq.(\ref{eq:sol}) into the definition of the Jarzynski work (for a time $\tau$), 
\begin{eqnarray}
W_J & = & -\int_0^\tau dt \dot{f}(t)[ H(t)x_o+G(t)v_o] \nonumber \\ 
& - &\int_0^\tau dt \int_0^t dt^\prime \dot{f}(t)G(t-t^\prime)
[f(t^\prime)+\eta(t^\prime)].
\end{eqnarray}
With the use of the
equilibrium averages, we find the mean of the Jarzynski work,
\begin{eqnarray}
<W_J>  & = & -\int_0^\tau dt \dot{f}(t) H(t) f(0)/\omega_o^2 \nonumber \\
& - & \int_0^\tau dt \int_0^t dt^\prime \dot{f}(t)G(t-t^\prime)f(t^\prime) .
\label{eq:meanjar}
\end{eqnarray}
We can re-express Eq.(\ref{eq:meanjar}) for later comparison with
$\sigma_J^2$ by integrating by parts and using the identity for
$\dot{H}(t)$ in Eq.(\ref{eq:ids}),
\begin{equation}
<W_J> = \Delta F+ \int_0^\tau dt \int_0^t dt^\prime \dot{f}(t)
H(t-t^\prime)\dot{f}(t^\prime)/\omega_o^2.
\label{eq:meanjar2}
\end{equation}

We find the variance of $W_J$ by using the oscillator equilibrium
averages and the fluctuation-dissipation theorem,
\begin{eqnarray}
\beta\sigma_{J}^2 & = & \left[\int_0^\tau dt \dot{f}(t)H(t)\right]^2
\frac{1}{\omega_o^2} + \left[\int_0^\tau dt \dot{f}(t)G(t)\right]^2 
\nonumber \\
& + & \int_0^\tau dt_1 \int_0^{t_1} dt_1^\prime \int_0^\tau dt_2 
\int_0^{t_2} dt_2^\prime \gamma(t_1^\prime-t_2^\prime) \nonumber \\
& \times & \dot{f}(t_1)G(t_1-t_1^\prime)\dot{f}(t_2)G(t_2-t_2^\prime).
\label{eq:varjar} 
\end{eqnarray}
In order to simplify this expression to compare with
Eq.(\ref{eq:meanjar2}), we must reduce the quadruple integral to a
more manageable double integral. We define
\begin{equation}
I(t_1,t_2) = \int_0^{t_1} dt_1^\prime \int_0^{t_2} dt_2^\prime 
G(t_1-t_1^\prime)G(t_2-t_2^\prime)\gamma(t_1^\prime-t_2^\prime).
\label{eq:conv}
\end{equation}
The integrals of $I(t_1,t_2)$ can be evaluated by first doing a double
Laplace transform, $\tilde{I}(s_1,s_2)=\int_0^\infty
dt_1 \int_0^\infty dt_2 e^{-s_1 t_1} e^{-s_2 t_2} I(t_1,t_2)$. This
double transform can be done by using the even symmetry of $\gamma(t)$.
We separate and define the two symmetric parts of $\gamma(t)$ using
the step function,
\begin{equation}
\gamma(t) = \gamma_+(t)+\gamma_-(t) = \gamma(t)\Theta(t)+\gamma(-t)\Theta(-t).
\end{equation}
Incidentally, the Laplace transforms of the two separate parts are
equal to the Laplace transform of $\gamma(t)$,
$\tilde{\gamma}_+(s)=\tilde{\gamma}_-(s)=\tilde{\gamma}(s)$. We use
this property, the convolution theorem, and Eqs.(\ref{eq:ids}) to
find the double transform,
\begin{equation}
\tilde{I}(s_1,s_2) = \frac{\tilde{H}(s_1)+\tilde{H}(s_2)}{\omega_o^2(s_1+s_2)} 
-\frac{\tilde{H}(s_1)\tilde{H}(s_2)}{\omega_o^2}-\tilde{G}(s_1)\tilde{G}(s_2).
\end{equation}
The inverse transform can easily be done, giving
\begin{equation}
I(t_1,t_2) = \frac{H(t_1-t_2)}{\omega_o^2}-\frac{H(t_1)H(t_2)}{\omega_o^2}
-G(t_1)G(t_2).
\end{equation}
Finally, we plug this expression back into Eq.(\ref{eq:varjar}),
\begin{eqnarray}
\beta\sigma_J^2 & = & \int_0^\tau dt \int_0^\tau dt^\prime \dot{f}(t)
H(t-t^\prime)\dot{f}(t^\prime)/\omega_o^2  \nonumber \\
& = & 2<W_J> - 2\Delta F.
\label{eq:varjar2}
\end{eqnarray}
The last equality proves the Jarzynski equality for harmonic
oscillators, even when $\eta(t)$ is not $\delta$-correlated. As in
Markovian stochastic derivations of the TFT~\cite{crooks,ND,polymer},
the generalized work $W_J$ does {\it not} satisfy the TFT, however
$W_{diss} = W_J-\Delta F$ does.

A simpler derivation of the JE follows if we assume time-translation
invariance of various correlation functions.
The definition of $\sigma_J^2$ contains the auto-correlation of
$\Delta x(t)=x(t)-<x(t)>$,
\begin{equation}
\sigma_J^2 = \int_0^\tau dt \int_0^\tau dt^\prime \dot{f}(t)
\dot{f}(t^\prime) <\Delta x(t)\Delta x(t^\prime)>.
\label{eq:varjar3}
\end{equation}
We evaluate the correlation in the brackets by using time-translation
invariance and the formal solution of $\Delta x(t) = H(t)\Delta
x_o+G(t)v_o + \int_0^t dt^\prime G(t-t^\prime)\eta(t^\prime)$,
\begin{equation}
<\Delta x(t)\Delta x(0)> = H(t)/(\beta\omega_o^2).
\end{equation}
Plugging this back into Eq.(\ref{eq:varjar3}), we recover the result
from the previous derivation Eq.(\ref{eq:varjar2}) and the JE is
easily seen.

A similar analysis is done for the real mechanical work,
\begin{eqnarray}
W & = & \int_0^\tau dt f(t)[\dot{H}(t)x_o+\dot{G}(t)v_o] \nonumber \\
& + & \int_0^\tau dt \int_0^t dt^\prime \dot{G}(t-t^\prime)
[f(t^\prime)+\eta(t^\prime)].
\end{eqnarray}
The average work is obtained using the equilibrium averages,
\begin{eqnarray}
<W> & = & \int_0^\tau dt f(t)\dot{H}(t)f(0)/\omega_o^2 \nonumber \\
& + & \int_0^\tau dt \int_0^t dt^\prime f(t) \dot{G}(t-t^\prime)f(t^\prime).
\label{eq:meanreal}
\end{eqnarray} 
The Laplace transform manipulations and the symmetric property of
$\gamma(t)$ can be used to find an expression for the variance
of the work $\sigma^2$. However, we only show the simpler
derivation, analogous to Eq.(\ref{eq:varjar3}),
\begin{equation}
\sigma^2 = \int_0^\tau dt \int_0^\tau dt^\prime f(t)f(t^\prime) 
<\Delta \dot{x}(t)\Delta \dot{x}(t^\prime)>,
\label{eq:varreal}
\end{equation}
where $\Delta \dot{x}(t) = \dot{x}(t)-<\dot{x}(t)>.$ The quantity
in brackets can be calculated by using the formal solution for the
velocity, i.e. the time derivative of Eq.(\ref{eq:sol}). Time-
translation invariance is again assumed and the velocity auto-correlation
is easily calculated
\begin{equation}
<\Delta \dot{x}(t)\Delta \dot{x}(0)> = \dot{G}(t)/\beta.
\end{equation}
We use this result in Eq.(\ref{eq:varreal}) and find
\begin{eqnarray}
\beta\sigma^2 & = & \int_0^\tau dt \int_0^\tau dt^\prime f(t)
\dot{G}(t-t^\prime)f(t^\prime) \nonumber \\
& = & 2<W>.
\label{eq:varreal2}
\end{eqnarray}
The last equality is from a comparison with Eq.(\ref{eq:meanreal})
and is valid when $f(0)=0$. Under this condition, we thus prove the
TFT for the probability distribution for the real mechanical work.

\begin{figure}
\begin{center}
\includegraphics[width=3.25in]{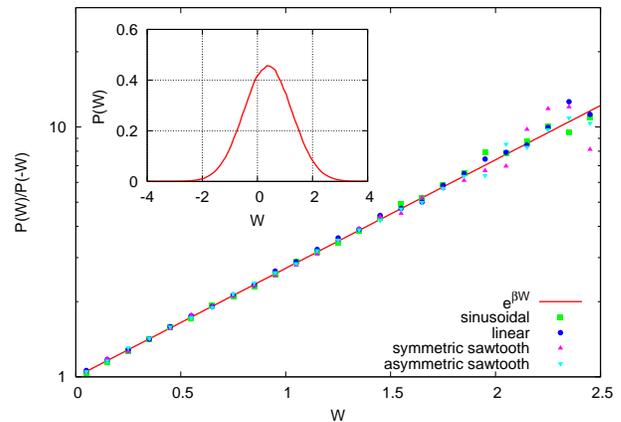}
\caption{(Color online) The inset shows the work probability
distribution $P(W)$ for a harmonic oscillator under a driving force
$f(t) = sin(\pi t/\tau)$ and a bath with exponentially correlated
noise. Simulation details are given in Sec. IV. The distribution
is a Gaussian with a positive mean, but with significant amplitudes
for negative $W$. Work probability distributions for other driving
forces are qualitatively similar, as is $P(W_{diss})$. The main
figure shows a log-lin plot of $P(W)/P(-W)$ versus $W$. The different
points are for a variety of driving forces and their excellent
agreement with the exponential confirms the validity of the TFT.}
\label{fig:harmonic}
\end{center}
\end{figure}

Eq.(\ref{eq:varjar2}) and Eq.(\ref{eq:varreal2}) are the main results
of this section. Simulations of a driven harmonic oscillator in a
non-Markovian bath confirm these derivations; for all driving forces
and bath conditions simulated, the JE and TFT are true for the
harmonic oscillator. Figure~\ref{fig:harmonic} shows the results
of these simulations. Details of the numerics and simulation results
for anharmonic oscillators are given in the next section.

\section{Anharmonic Oscillators}

For anharmonic oscillators, finding simple expressions for $P(W)$
and $P(W_J)$ is, in general, not possible; $W$ and $W_J$ are not
linear combinations of the Gaussian stochastic quantities, therefore
their distributions are not Gaussian. In this section, we numerically
measure the work distribution functions for oscillators with the
potential of Eq.(\ref{eq:pot}) when $k_3,k_4\geq 0$.

Random uncorrelated Gaussian numbers are commonly
numerically generated and used in standard integration algorithms
for Langevin equations of motion. An efficient Verlet-like integrator
for second order white noise Langevin equations is given in
Ref.~\cite{verlet}. We modify this algorithm in order to simulate
Langevin systems with exponentially correlated noise. The damping
kernel has the form
\begin{equation}
\gamma(t) = e^{-\Gamma|t|}.
\label{eq:expcorr}
\end{equation}

Exponentially correlated noise can be effectively introduced by
using white noise in the equation of motion for an auxiliary variable
$z$, in addition to the position $x$ and velocity $v$. The coupled
differential equations are
\begin{eqnarray}
dx/dt & = & v, \nonumber \\
dv/dt & = & -dV(x)/dx + f(t) + z, \nonumber \\
dz/dt & = & -\Gamma z - v + \zeta.
\label{eq:eom}
\end{eqnarray}
$\zeta$ represents Gaussian white noise which is easily generated
numerically,
\begin{equation}
<\zeta(t)>=0, \qquad <\zeta(t)\zeta(t^\prime)> = 2T\delta(t-t^\prime).
\end{equation}
It can be shown that the equations of motion Eqs.(\ref{eq:eom})
are equivalent to Eq.(\ref{eq:langevin}) with exponentially correlated
noise Eq.(\ref{eq:expcorr}). In all simulations, $T=\Gamma=\omega_o^2=1$,
though we have verified that our results are qualitatively the same
with different temperatures and damping constants. We use a step
size $h$ with $h=0.01$ in all simulations shown. We have checked
that our results do not change with smaller step sizes.

Work distributions are approximated by histograms of $O(10^6)$
measurements of the work done ($W$ and $W_J$) over a time $\tau$.
Between each measurement, we allow the oscillator to equilibrate
by integrating Eqs.(\ref{eq:eom}) for $O(10^5)$ steps with $f(t)=0$.
No differences in the distributions are seen for different histogram
bin sizes and longer equilibration times.

We use a sawtooth driving force of the form,
\begin{eqnarray}
f(t) & = & t/t_o, \qquad 0 \leq t \leq t_o \nonumber \\
& = & (-t+\tau)/(\tau-t_o), \qquad t_o < t \leq \tau.
\label{eq:saw}
\end{eqnarray}
Under this driving force, the equilibrium configurations at $t=0$
and $t=\tau$ are identical, i.e. $\Delta F=0$ and $W=W_J$. In the
$\tau\rightarrow 0$ limit, the work distribution is trivially peaked
at $W=0$ because there is no time for work to be done. $P(W)$ is
similarly peaked at zero in the $\tau\rightarrow \infty$ limit
because the system is driven adiabatically from an equilibrium state
back to the same equilibrium state. Our numerical simulations are
in accord with these physical limits. The figures show simulation
results with $\tau=10$, which is intermediate between these two
limits. (No qualitative differences in terms of the JE, TFT, and
CT exist with different $\tau$.) Lastly, $f(0)=0$, which is a
necessary condition for the validity of the TFT for the harmonic
oscillator detailed in the last section and Figure~\ref{fig:harmonic}.

By changing $t_o$, we can alter the symmetry of $f(t)$; the force
is symmetric in time only when $t_o=0.5\tau$. We implement this
symmetric driving and two asymmetric sawtooth forces with
$t_o=0.25\tau,0.75\tau$ in our simulations. As mentioned in the
introduction, there is a subtle difference in the TFT and the CT
with the latter using a time-reversed process in the denominator
of the ratio of probabilities. For the symmetric sawtooth force,
forward and reverse driving must be equivalent and no differences
exist between the TFT and the CT. Our asymmetric sawtooth forces
are complementary in that the time-reverse driving of one force is
equivalent to the time-forward driving of the other. In other words,
in terms of this sawtooth force, the TFT is
\begin{equation}
\frac{P_{t_o}(+W)}{P_{t_o}(-W)} = e^{+\beta W},
\label{eq:TFTsaw}
\end{equation}
whereas the CT is
\begin{equation}
\frac{P_{t_o}(+W)}{P_{\tau-t_o}(-W)} = e^{+\beta W}.
\label{eq:CTsaw}
\end{equation}
($P_s(W)$ is the work distribution corresponding to a saw-tooth
potential with a break at time $t=s$.) 

\begin{figure}
\begin{center}
\includegraphics[width=3.25in]{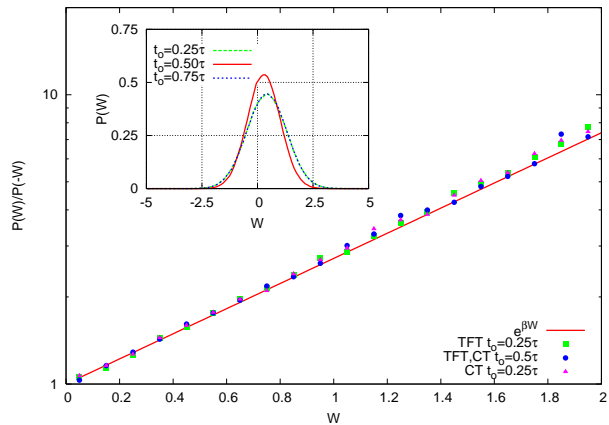}
\caption{(Color online) The inset shows the work probability
distributions for oscillators with quartic anharmonicities under
sawtooth driving with $t_o/\tau=0.25,0.5,$ and $0.75$. These
distributions have positive means, but also substantial weight on
the negative region of $W$. $P_{0.25\tau}(W)=P_{0.75\tau}(W)$,
indicating that the asymmetry in the driving force does not alter
the work distribution functions, thus the ratios in the definition
of the TFT and the CT, Eq.(\ref{eq:TFTsaw}) and Eq.(\ref{eq:CTsaw}),
must also be equal. The main figure confirms this equality and shows
that the TFT and the CT are valid for the quartic oscillator.}
\label{fig:beta}
\end{center}
\end{figure}
We first simulate an anharmonic spring with a unit quartic anharmonicity
(i.e. $k_3=0, k_4=1$). Figure~\ref{fig:beta} displays the results
of these simulations. Aside from not being Gaussian, the probability
distributions have similar properties as the work distribution of
the harmonic oscillator: the means are all positive and there is a
very substantial probability of measuring negative work. Furthermore,
the two complementary asymmetric driving forces give the same work
probability functions. From Eq.(\ref{eq:TFTsaw}) and Eq.(\ref{eq:CTsaw}),
it is clear that, in this case, the TFT and the CT are equivalent.
The main plot in Fig.~\ref{fig:beta} displays this equality as well
as the validity of the theorems.

\begin{figure}
\begin{center}
\includegraphics[width=3.25in]{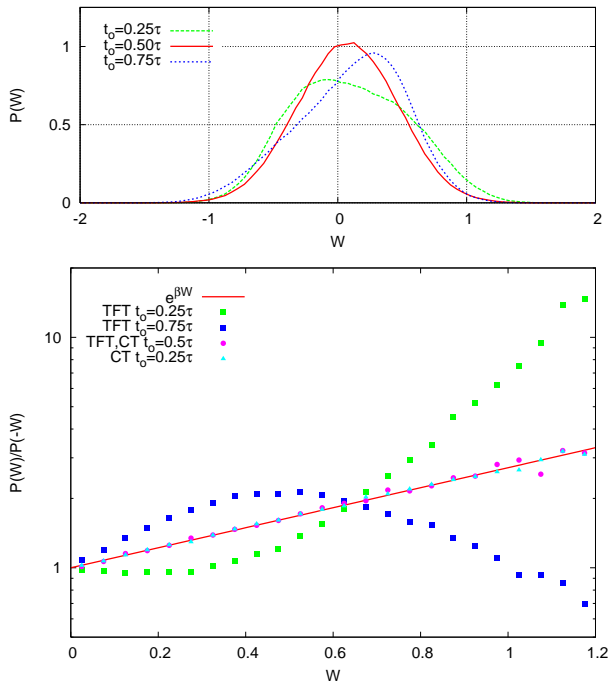}
\caption{(Color online) The top panel shows the work probability
distributions for particles in an asymmetric potential under sawtooth
driving with $t_o/\tau=0.25,0.5,$ and $0.75$. Due to the lack of
symmetry in the potential, $P_{0.25\tau}(W)\neq P_{0.75\tau}(W)$.
The bottom panel shows the ratios from Eq.(\ref{eq:TFTsaw}) and
Eq.(\ref{eq:CTsaw}) versus $W$ on a log-lin plot. The data sets
that agree with the exponential indicate that the CT is valid even
for asymmetric potentials. The other sets indicate that the TFT
fails for these potentials under general driving forces. }
\label{fig:cube}
\end{center}
\end{figure}
When a cubic nonlinearity is included ($k_3=k_4=1$), the even symmetry of
the potential is broken. Figure~\ref{fig:cube} displays results for
these non-symmetric oscillators. The top plot is analogous to the
inset of Fig.~\ref{fig:beta} and shows the probability distributions
for the three different sawtooth driving forces. From this figure,
it is clear that all three distributions are different and the ratio
in the TFT Eq.(\ref{eq:TFTsaw}) is not equivalent
to the ratio in the CT Eq.(\ref{eq:CTsaw}).

The difference between the TFT and the CT is clearly shown by the
bottom panel of Fig.~\ref{fig:cube}. In fact, the CT is valid under
all sawtooth driving forces, whereas the TFT is only valid in the
symmetric driving force. For oscillator potentials that are not
symmetric in $x$, i.e. $V(x)\neq V(-x)$, and the driving force is
asymmetric in time about $\tau/2$, the TFT dramatically fails.

For our example the validity of the JE follows from the CT.  
Independently, we have also measured the average exponential of the 
work and find 
$<e^{-\beta W}>=1$ within $\sim 4$ decimal places for both symmetric
and asymmetric oscillators. Because $\Delta F=0$, this indicates
the Jarzynski equality Eq.(\ref{eq:JE}) is most likely valid for a
large class of oscillator potentials, even with exponentially
correlated noise.

When a sinusoidal force of the form $f(t)=sin(n\pi t/\tau)$, where
$n$ is an odd integer, is used, the simulation results are qualitatively
the same as with the symmetric sawtooth force. Obviously, this is
because the period of the sinusoidal force gives a symmetric (about
$\tau/2$) driving process. We also simulate linear driving which
takes the system to a different equilibrium configuration in the
adiabatic limit. Though $\Delta F \neq 0$ with linear driving, we
find that the JE and the CT are still valid, but the TFT fails.
This last result is not surprising because the linear driving force
has many qualitative similarities with asymmetric sawtooth driving
forces.

The above results are unchanged when colored noise is replaced
by white noise. The white noise results are to be expected due to
the numerous analytical derivations of these nonequilibrium
theorems~\cite{jarzynski,crooks,ND,markov1,markov2,markov3} for
Markovian baths. Our simulation results shown in this section
indicate that the CT and the JE are also valid for non-Markovian
baths.

\section{Summary}

We have derived non-equilibrium work fluctuation theorems for the
classical harmonic oscillator connected to a generalized Langevin bath by
studying the work distribution functions of the real mechanical
work and the Jarzynski generalized work. Both  $W$ and $W_J$ are
Gaussian variables. We derive the TFT for the real work and the JE
by using exact relations between $<W>$ and $\sigma^2$ and $<W_J>$
and $\sigma_J^2$. To our knowledge, these are the first rigorous derivations 
of the nonequilbrium work fluctuation theorems for a system described by an
arbitrary damping kernel of the generalized Langevin equation.

We also numerically measure the work distribution functions for
anharmonic oscillators in Langevin baths with exponentially correlated
noise. Our numerical simulations indicate that the JE extends to
particles in a variety of oscillator potentials and driving forces.
They also show that the symmetries of the potential and the driving
force account for differences between the TFT and the CT. Crooks'
theorem is valid for all potentials and driving forces simulated,
whereas the TFT is not valid when the oscillator potential is not
even in $x$ {\it and} the driving force is not symmetric in $t$
(about $\tau/2$). The strong agreement between our simulation results
and the nonequilibrium theorems (CT and JE) provides motivation for
a derivation of the theorems for anharmonic oscillators in non-Markovian
baths.

These results are important due to the existence of noise correlations
in any real heat bath. This has obvious consequences for experimental
tests of the Jarzynski equality and the fluctuation theorems.
Experimentally, the JE is a powerful tool to measure equilibrium
free energy differences due to the fact that any real driving is
done irreversibly. Interesting and open questions still exist on
the ramifications of the nonequilibrium fluctuation theorems and
the second law ``violating'' events for molecular engines and
microscopic thermodynamics.

\end{document}